# An all optical broadband tunable quantum frequency shifter

Li Chen[1, 2,#], Zhi-Yuan Zhou[1,2,#,*], Ming-Yuan Gao[1 2], Wu-Zhen Li[1 2], Zhao-Qi-Zhi Han[1 2], Yue-Wei Song[1 2], Ren-Hui Chen[1 2], and Bao-Sen Shi[1,2,*]

[1] *CAS Key Laboratory of Quantum Information, University of Science and Technology of China, Hefei, Anhui 230026, China*

[2] *CAS Center for Excellence in Quantum Information and Quantum Physics, University of Science and Technology of China, Hefei 230026, China*

[#]*These two authors contributed equally to this article.*

\* *zyzhouphy@ustc.edu.cn*

\* *drshi@ustc.edu.cn*

A frequency shifter of the photon is a key component for frequency-multiplexed high-capacity quantum communications and frequency-encoded quantum computation. Existed methods for shifting the frequency of a photon based on electro-optical, or acousto-optical effect, however, suffer the limited frequency shift up to a few hundreds of GHz, furthermore, high-quality micro-wave electronics are required. The frequency of a photon can also be shifted with the frequency difference equal to the frequency of pump laser by using an all optical-wave-mixing approach, which is usually about tens of THz. So, there is a big frequency shifting gap between these methods. Here, we propose a new scheme of a quantum frequency shifter based on an all-optical wave-mixing process, which can theoretically achieve a frequency shift ranging from GHz to a few THz, therefore bridging the gap. As a principle of poof, by using two pump beams in a three-wave mixing cascading process, a heralded single photon is frequency-shifted more than ±400GHz, and the shift can be tuned continuously over broadband by changing the frequency difference between two pump lasers. Besides, high coincidence to accidence ratio between the shifted photons and the heralded photon indicates the preserve of quantum properties. The present quantum frequency shifter is in analog to an electro-optical based shifter, but with much broader tuning ability. Our all-optical quantum frequency shifter will become a fundamental building block for high-speed quantum communication networks and frequency domain photonic quantum computation.

A photon is a basic information carrier in quantum information science and technology due to its weak coupling to the environment. Basic degree of freedoms of a photon includes: path, position, polarization, orbital angular momentum and frequency, etc. The path, polarization and orbital angular momentum can be readily well controlled with linear optical elements such as beam splitters, waveplates and holographic gratings [1, 2, 3], on the contrary, manipulating frequency of a photon is much more complex and a big challenge since it alters the energy of photon. Electro-optical[4, 5, 6, 7, 8], acousto-optical[9, 10, 11, 12] effects and an all-optical wave mixing process [13, 14, 15, 16, 17, 18, 19] are typical approaches used to change the frequency of a photon. In quantum information, to shift the frequency of a photon is a basic requirement for frequency-multiplexed high-capacity quantum communication[20, 21, 22] and frequency-domain quantum computation[23, 24, 25]. Significant progresses about shifting the frequency of a light have been achieved: with electro-optical effect, by using coupling integrated micro-cavities, nearly 30-GHz bi-directional frequency shift is obtained for a single shift process, and nearly 120-GHz frequency shift is achieved by using cascaded processes[7]; Frequency shifting of quantum light with cascaded electro-optical modulators is also realized, the maximum frequency shift reaches ±200 GHz [6]; With opto-mechanical effect, a high efficient and noiseless frequency shift up to 150 GHz is realized [26]; By using all-optical wave mixing, large frequency shifting over tens of THz has been realized with high efficiency [13, 14, 15, 16, 17, 18, 19, 27]. Though electro-optical and opto-mechanical based frequency shifters are promising, their tuning ability is limited to a few hundreds of GHz, besides, high quality micro-wave electronics are required. Additionally, the present all optical-wave-mixing approach usually shifts photon frequency to tens of THz. Obviously, there is a big frequency shifting gap between these existed methods, so how to bridge this gap is an open question.

Here we propose a new all-optical wave-mixing approach that can well solve the above questions. In traditional three wave mixing schemes[13, 14, 15, 16, 17, 18, 19, 27], only one pump laser exists, therefore the frequency of a photon is only shifted with the frequency difference equal to the frequency of pump laser. In our scheme, by using two pump laser beams, we show that the frequency of the photon can be shifted symmetrically with the frequency difference between the two pump laser beams. This new frequency shifter is in analog to an electro-optical based modulator, but with a broader tuning bandwidth determined by the phase matching condition. More interesting is that by using some special nonlinear crystal, for example, a chirped periodic poled waveguide, the phase-matching bandwidth can be enlarged significantly, so this all-optical quantum wavelength mixing approach can theoretically achieve a frequency shifting up to a few THz. In a principle of proof experiment, by using a chirped poling LN (CPPLN) crystal, we continuously shift the frequency of a heralded single photon more than ±400 GHz, covering more than eight 100-GHz dense wave multiplexing channels. Coincidences to accidences ratio (CAR) of more than 30 between the shifted photon and the heralding photon indicates the preservation of the nonclassical properties between photons. This work opens up a new avenue for building an all-optical broadband tuning quantum frequency shifter, which will function as a basic component in building quantum networks.

Firstly, we will introduce the general method of our frequency shifter, the whole scheme consists of multiple cascaded nonlinear processes. As shown in Fig. 1(a-b), the photon frequency conversion process based on the first-order cascaded process in a CPPLN crystal can be divided into two steps. The initial input lights consist of a 1550-nm signal photon S and two continuous-wave (CW) 1064 nm pump lasers P1 and P2 with slightly different wavelength. At the first step shown in Fig. 1(a), two visible band photons SF1 and SF2 are generated through sum frequency generation (SFG) between the signal photon S and the pump lasers P1, P2 (these two SFG processes are named SFG1 and SFG2, respectively); At the second step, the first-order cascading photons DF1 and DF2 are generated by the difference frequency generation (DFG) between SF2, SF1 and P1, P2 lasers (these two DFG processes are named DFG1 and DFG2). At this point, the photons DF1, DF2 and the signal photons S still maintain the quantum correlation properties, and the frequency of the converted photons DF1, DF2 can be continuously shifted. As shown in Fig. 1(c-d), a high-order cascaded process can generate high-order cascaded photons with frequencies near those of the signal photons, which can realize broader bandwidth photon frequency shifting as well as the generation of a single photon with multi-frequency components. Due to the low conversion efficiency of the cascaded processes in our experimental system, the second-order conversion efficiency is 2-3 orders of magnitude lower than the first-order processes, therefore we currently only consider the first-order processes as shown in Fig. 1(a-b). However, if the pump power and conversion bandwidth can be increased, the higher-order cascading photons can be generated efficiently in sequential cascaded processes, which functions as an effective way to establish quantum entanglement networks with tunable frequency interval and multi-channel multiplexing.

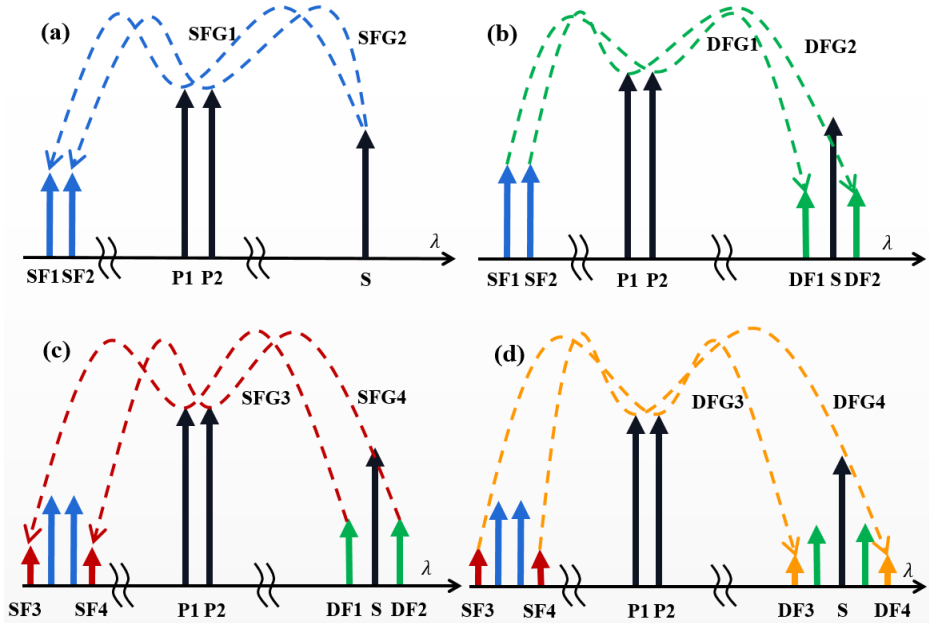

Fig. 1. Schematic diagram of the four steps (a-d) in the quantum cascading processes.

The schematic diagram of the experimental setup is shown in Fig. 2. The experimental system consists of three modules: SPDC part, cascading part and coincidence part. In the SPDC part, 1550 nm correlated photon pairs with high signal-to-noise ratio are generated via the spontaneously parametric down-conversion (SPDC) process in a bulk PPKTP crystal. The used type-II PPKTP (Raicol crystals) crystal has dimension of 1 mm × 2 mm × 15 mm, with a periodical poling period of 46.2 μm. The pump laser for SPDC process comes from a diode laser (TOPTICA pro, Graefelfing), whose spatial mode is optimized by passing it through a single-mode fiber. The pump laser was set to vertical polarization by waveplates before the nonlinear crystal. The temperature of the crystal was controlled by using a home-made temperature controller with a temperature stability of ±2 mK. The two generated photons with different polarizations are separated by a PBS and coupled into the optical fibers after passing through a long-pass filter, respectively. One of the photons acts as the reference photon, passing through two 100 G fiber filters with a center of C34 (correspond to the standard International Telecommunications Union (ITU)), and the other photons are connected to the cascading module as the signal photon through two same filters.

In the cascading module, we use a type-0 phase-matched CPPLN crystal with poling period of 11.75-11.85μm to construct the cascading conversion system. The wavelength of laser P1 is fixed at 1063.9 nm, and the wavelength of laser P2 can be continuously tuned near 1063.9 nm. Two seed lights with small wavelength interval are combined by a fiber beam splitter and amplified by the $Yb^{3+}$ doped fiber amplifier. The pump lights overlap with the signal photon after a dichroic mirror DM (with high transmission at 1064 nm and high reflection at 1550 nm) and then focused by a lens set with beam waists of 50 μm and 60 μm, respectively. The photons produced by the cascading process have a frequency shift with respect to the signal photons, and the shift depends on the wavelength difference between the two 1064 seed lights. Subsequently, the cascading photons are coupled into an optical fiber after passing through a long-pass filter and a band-pass filter. In the coincidence measurement part, the cascading photons passing through 100-GHz fiber filters of different channels (C30-C38) are input into single-photon detectors. The frequency-shifted photons and the reference photons are detected using two superconducting nanowire single-photon detectors (SNSPD1 and SNSPD2, with detection efficiencies of 80% and dead times of less than 100 ns). Finally, the signals from the SNSPDs are sent to a counting board (PicoQuant TimeHarp 260) to record coincidence events with a 0.2 ns time bin.

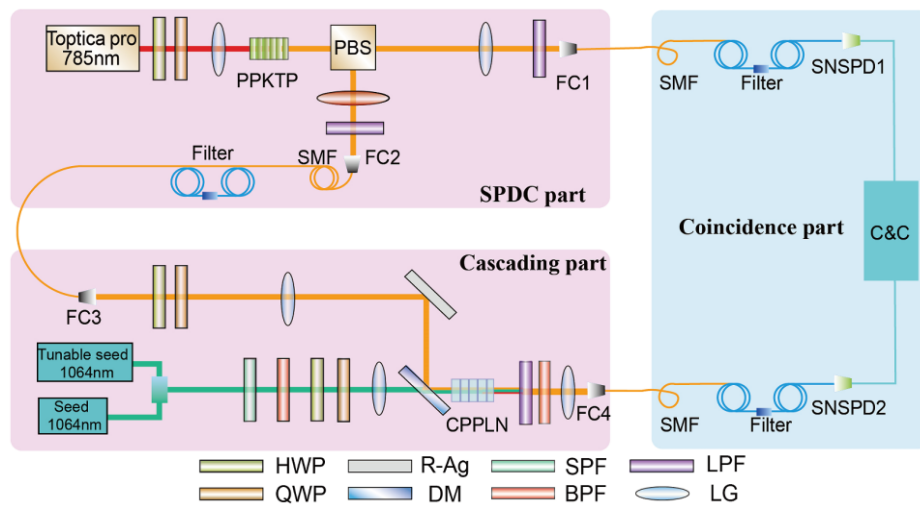

Fig. 2 Schematic illustration of the experimental setup for the quantum frequency shifter system. PPKTP: periodically poled potassium phosphate titanate that has size 15 mm*2 mm*1mm (height); CPPLN: periodically poled lithium niobate crystal that has size 40 mm*3 mm*1 mm (height); YDFA: $Yb^{3+}$ doped fiber amplifier; HWP: half wave plate; QWP: quarter wave plate; LG: lens group; R-Ag: Ag mirrors;

PBS: polarizing beam splitter at 1550 nm; DM: 1064T/1550R dichroic mirror; LPF:1400 nm long-pass filter; BPF: 1075-50nm (or 1550-12nm) band-pass filters; FC1-FC4:fiber couplers; SMF: single mode fiber at 1550nm; SNSPD1(SNSPD2): superconducting nanowire single-photon detector; Filter: 100Ghz fiber filters of different channels(C30-C38).

Firstly, we use the SFG process in the CPPLN as an example to characterize the nonlinear optical performance of the CPPLN crystal. Before it reached the CPPLN crystal, the power of the pump laser was 10 W, and the signal laser power was 50 mW. A 6.1 mW visible laser beam at 633 nm was obtained via SFG process. As shown in Fig. 3(a), the output SFG laser power is directly proportional to the pump power due to the low conversion efficiency. Taking the 6.68% power loss caused by the subsequent filters and the dichroic mirrors into account, the power efficiency of the SFG process defined by $\eta_{power} = P_{633}/P_{1550}$ was 12.2%, and the corresponding quantum conversion efficiency (QCE), as defined by $\eta_{quantum} = \eta_{power} \lambda_{633}/\lambda_{1550}$, was 4.98%. Furthermore, by setting the temperature of the crystal to be at 30°C, and varying the wavelength of the signal light to measure the normalized QCE (NQE), we obtained the 0.8 nm conversion bandwidth of the SFG process at 1064 nm. The wavelength-temperature adjustment coefficient of the PPLN crystal was measured to be 0.151 nm/°C at 1064 nm, and the measured crystal characteristic curves were as shown in Fig. 3(b) and 3(c).

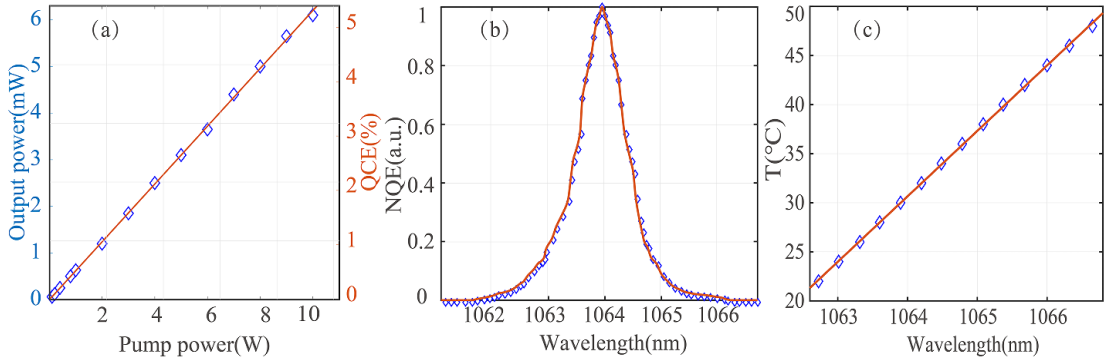

Fig. 3 (a): Relationships between pump power and SFG output power quantum efficiency in CPPLN; (b)-(c): wavelength and temperature tuning characteristics of SFG in CPPLN.

Next, we characterize the two-photon source. An important parameter is the CAR, which is defined as the ratio of the coincidence counts to the accidental coincidence counts. The CAR values and coincidence counts are given for different pump powers in Fig. 4(a), with an integration time of 60 s and a coincidence window of 0.2 ns. The CAR firstly increases monotonically with the increase of the pump power, then rapidly decreases when beyond a certain threshold, this is due to the growing multiphoton emission in the SPDC process. Afterward, we performed coincidence measurement of the frequency shifted photons with the reference photons. The 1064 nm pump power for the cascading process is 16 W and the pump power for the SPDC process is 20 mW, the CAR of the corresponding channels (C33-C35,C32-C36,C31-C37,C30-38) with different frequency shifts (100 GHz, 200 GHz, 300 GHz, 400 GHz) were tested by adjusting the wavelength spacing of the two 1064 seed lights and matching the corresponding 100 G filters as shown in Fig. 4(b), and the normalized transmittance curve of the filters for different channels (C30-C38,C31-C37,C32-C36,C33-35) in the experimental system are shown in Fig. 4(c), we find that with the increase of frequency shift, the CAR decreases with the decrease of the conversion efficiency of the cascading process.

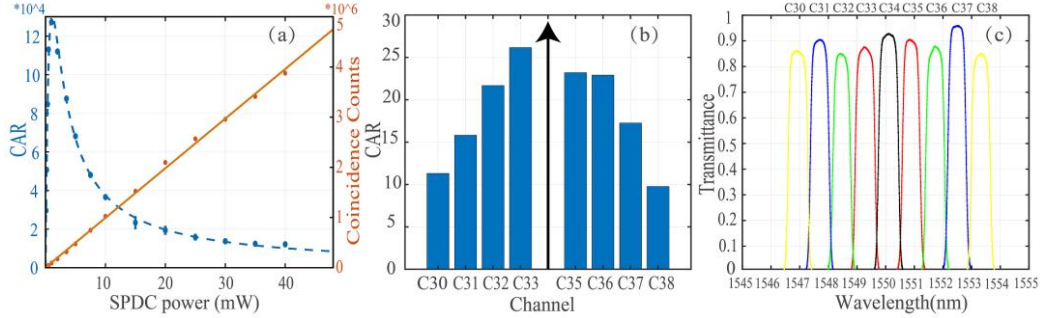

Fig. 4 (a): Coincidence counts and CAR curve with varying 775 nm SPDC pump power of SPDC process in PPKTP crystals, and the single measurement time is 60 s. (b): the CAR of the corresponding channels (C30-C38, C31-C37, C32-C36, C33-35) at different frequency shifts (100 GHz, 200 GHz, 300 GHz, 400 GHz) in the coincidence measurements of the frequency shifted photons after cascading process and the reference photons. The pump power in the SPDC process is 20mW, the pump power in the cascading process is 16W, and the single measurement time is 60 s. (c): Measured transmittance of 100Ghz filters with different channels (C30-C38, C31-C37,C32-C36,C33-35).

Finally, we tested the performance characteristics of the quantum frequency shifter based on the cascading process. Fig. 5(a) gives the CAR values of coincidence measurements between the frequency shifted photons with the reference photons at different SPDC pump power, the CAR firstly increases monotonically with the increase of the pump power, then gradually stabilizes beyond a certain threshold. We will provide a detailed derivation of the principles behind this relationship in the Methods section. And Fig. 5(b) shows the CAR at different pump powers in the cascading process, and the curve of CAR versus pump power is basically linear, this is due to the cascading process contains two second-order nonlinear processes, the conversion efficiency and coincidence counts exhibit a quadratic relationship with cascading pump power, while the accidental coincidence count is related to the Raman noise of the pump light in the CPPLN crystal, and the noise is linear against the pump power, thus the relationship between the CAR and the cascading pump power is linear as shown in Fig. 5(b). The analysis of the coincidence counts and the accidental coincidence counts of the quantum frequency shifter system is shown in the Supplementary Information. Additionally, we also tested the relationship between CAR and the temperature of the CPPLN crystal in the cascading system with the same pump power as shown in Fig. 5(c), and it basically satisfies the $\sin c^4$ function, please refer to the Method for further details of the theoretical analysis about the Fig. 5(a)-(c). Finally, we tested the coincidence counts and accidental coincidence counts with no cascading pump light as well as with no signal photons at different pump power, respectively, and we found the net coincidence counts (it is defined by the coincidence counts minus accidental coincidence counts) is equal to zero in the whole test, please refer to the Supplementary Information for further details. As a result, the fluctuation of the pump light has a negligible effect on the coincidence measurements, and the isolation of the filter is high enough to make the leakage of center signal photons (C34) to other channels negligible too.

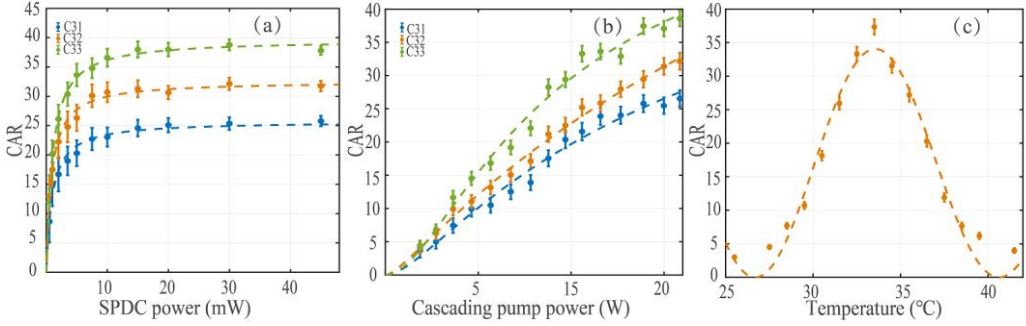

Fig. 5: Coincidence measurement of frequency shifted photons after cascading process and reference photons. (a) CAR curve with varying 775nm SPDC pump power. The pump power in the cascading progress is 16W, and the single measurement time is 60s. (b) CAR curve with varying cascading pump power at 1064 nm. The pump power in the SPDC process is 20mW, and the single measurement time is 60s. (c) CAR curve with varying the temperature of the CPPLN crystals in the cascading process. The pump power used for the SPDC process is 20mW and the pump power used in the cascading process is 16W, and the single measurement time is 60s.

Although the above proof-of-principle experiment of the all-optical wave-mixing approach can only continuously frequency tune beyond ±400 GHz, in principle, the broader tuning bandwidth of this approach depends only on the conversion efficiency and the phase-matching conditions of the cascading process. We can use wider chirped periodic poled waveguide to further increase the conversion bandwidth and the efficiency of the cascading process, and the phase-matching bandwidth of the common CPPLN waveguide can be up to a few THz[28]. Therefore, if a suitable CPPLN waveguide structure is used, this all-optical quantum wavelength mixing approach can theoretically achieve a frequency shifting up to a few THz. In addition, increasing the conversion bandwidth and efficiency of the cascading process not only increases the frequency shifting range of the first-order cascading process, but also promotes the efficient generation of higher-order cascade-converted photons, which can be seen from the schematic diagram of the principle that the tuning range of second-order cascading process is twice that of the first-order process. This advancement extends the continuous tuning range of the quantum frequency shifter, and enables the preparation of quantum light fields with multiple frequency components.

In summary, we propose and successfully demonstrated in principle an approach for manipulating frequency of a photon through a cascading second-order nonlinear process, leading to the development of an advanced quantum frequency shifter. By precisely controlling the wavelength difference between two pump lights in the cascading process, we have realized continuous frequency tuning over ±400 GHz, and maintained quantum coherence and avoided additional noise. By using some special nonlinear crystal like a chirped periodic poled waveguide, this all-optical quantum wavelength mixing approach can theoretically achieve a frequency shifting up to a few THz. Our method provides an efficient, broadband and clear solution for frequency control. The all-optical quantum frequency shifter represents a significant step forward in the development of high-capacity quantum entanglement networks featuring tunable frequency intervals and multi-channel multiplexing capabilities. This technology is poised to become an essential component in future high-speed quantum communication networks and frequency domain photonic quantum computation.

**Acknowledgments**
We would like to acknowledge the support from the National Key Research and Development Program of China (2022YFB3903102，2022YFB3607700), National Natural Science Foundation of China (NSFC)(62435018), Innovation Program for Quantum Science and Technology (2021ZD0301100), USTC Research Funds of the Double First-Class Initiative（YD2030002023）, and Research Cooperation Fund of SAST, CASC (SAST2022-075).

**Author contributions**
Z.-Y.Z. and B.-S.S. designed the experiment. L. C. carried out the experiment. Z.-Q.-Z H., M.-Y. G., W.-Z. L. and R.-H. C. helped collect the data. Z.-Y.Z. and L. C. analyzed the data and wrote the paper with help from all other authors. The project was supervised by Z.-Y.Z. and B.-S.S. All authors discussed the experimental procedures and results.

**Methods:**
**1. Details of the quantum theory of the quantum frequency shifter**
We describe the quantum theory of the quantum frequency shifter, mainly comprising the quantum theory for first-order cascading processes of continuous waves in second-order nonlinear crystals. Four waves are involved in the cascading process: two strong pump beam at frequency $\omega_{p1}, \omega_{p2}$, one signal beam needed to be converted at frequency $\omega_s$, one SFG beam with a transit role at frequency $\omega_{sfg}$ and the converted beam at frequency $\omega_c$, where the frequencies of the interacting waves satisfy the energy conservation $\omega_s + \omega_{p1} = \omega_{sfg}, \omega_{sfg} - \omega_{p2} = \omega_c$, and $i = s, p1, p2, sfg, c$ denotes the signal, pump, SFG photon frequencies and first-order cascading photon, respectively. Under the condition of perfect phase matching and undepleted pump approximation, we discuss the SFG and DFG processes separately, the associated Hamiltonian of the three-wave mixing can be written as [29]:

$$\begin{aligned}\hat{H}_{sfg} &= i\hbar g_1 E_{p1} \left( \hat{a}_s \hat{a}_{sfg}^\dagger - H.c. \right) \\ \hat{H}_{dfg} &= i\hbar g_2 E_{p2} \left( \hat{a}_{sfg} \hat{a}_c^\dagger - H.c. \right)\end{aligned} \qquad (1)$$

Here, $g_1, g_2$ is a constant determined by the second-order polarization tensor; $E_{p1}, E_{p2}$ is the electric field amplitude of the pump light; and $\hat{a}_s, \hat{a}_{sfg}, \hat{a}_c$ are the annihilation operators of the signal photon, the SFG photon and the cascading converted photon, respectively. By solving the Heisenberg equation, the analytical results of the signal and upconverted photon states are expressed as:

$$\hat{a}_{sfg}(L) = sin(g_1 E_{p1} L)\hat{a}_s(0) + cos(g_1 E_{p1} L)\hat{a}_{sfg}(0)$$
$$\hat{a}_s(L) = -sin(g_1 E_{p1} L)\hat{a}_{sfg}(0) + cos(g_1 E_{p1} L)\hat{a}_s(0) \quad (2)$$
$$\hat{a}_c(L) = sin(g_2 E_{p2} L)\hat{a}'_{sfg}(0) + cos(g_2 E_{p2} L)\hat{a}_c(0)$$

Here, L is the length of the crystal. The quantum efficiency of cascaded conversion is defined as $\eta = N_c(L)/N_s(0)$, where $N_i = \langle \hat{a}_i^\dagger \hat{a}_i \rangle$ is the average number of photons measured in a certain time. At the input, the number of cascaded converted photons $N_c(0) = 0$, and considering the result of the first step of the SFG as the initial condition for the DFG: $N'_{sfg}(0) = \frac{1}{2} N_{sfg}(L)$, When the interacting waves are all Gaussian modes, the cascaded conversion quantum efficiency can be written as $\eta = \frac{1}{2}\sin^2(\frac{\pi}{2}\sqrt{\frac{P}{P_{max1}}})\sin^2(\frac{\pi}{2}\sqrt{\frac{P}{P_{max2}}})$ [30], where P is the actual pump power and $P_{max1}, P_{max2}$ are the pump optical powers required to reach maximum quantum conversion efficiency for the SFG and DFG processes, respectively, $P_{max1}, P_{max2}$ expressed as follows[30, 31, 32]:

$$P_{max1} = \frac{\varepsilon_0 c n_s n_{sfg} \lambda_p \lambda_{sfg} \lambda_s}{16\pi^2 d_{eff}^2 L}$$
$$P_{max2} = \frac{\varepsilon_0 c n_c n_{sfg} \lambda_p \lambda_{sfg} \lambda_c}{16\pi^2 d_{eff}^2 L} \quad (4)$$

Here, $n_i (i = s, c, sfg)$ are the refractive indices of the beams inside the crystal, $\lambda_i (i = s, c, sfg)$ are the wavelength of the lights, $d_{eff}$ is the effective nonlinear coefficient, c is the light velocity, and $\varepsilon_0$ is the permittivity of a vacuum.

## 2. Details of the theoretical analysis about the coincidence measurement results between the frequency shifted photons and reference photons.

We conducted a brief theoretical analysis of the coincidence measurement between the frequency shifted photons and the reference photons to qualitatively explain the physical significance of the line shapes obtained in Figures 4.8 and 4.9. In the SPDC process, a pump photon is annihilated to generate a pair of correlated photons. The number of generated photons is proportional to the pump power and can be expressed as:

$$N_0 = aP_1 \tag{5}$$

Here, $N_0$ represents the number of photon pairs generated by the SPDC process, and $P_1$ denotes the pump power of the SPDC process. Taking into account the losses during photon transmission and the collection efficiency of the detectors, the actual detected photon count can be expressed as:

$$N_1 = \alpha_1 \eta_1 N_0 + \beta_1 P_1 + N_{n1} \tag{6}$$

$$N_2 = \alpha_2 \eta_2 N_0 + \beta_2 P_1 + N_{n2} \tag{7}$$

Where $\alpha_{1,2}$ is the transmittance of the two channels, $\eta_{1,2}$ is the detection efficiency, and $\beta_{1,2}$ represents noise photons unrelated to the parametric process, $N_{n1,2}$ represents the environmental noise and dark counts. In this case, the CAR can be expressed as:

$$CAR_{SPDC} = \frac{\alpha_1 \eta_1 \alpha_2 \eta_2 N_0}{N_1 N_2 \Delta \tau} \tag{8}$$

By substituting equations (5), (6), and (7) into the equation (8), the CAR can be expressed in the form of $\dfrac{1}{a_1 P_1 + b_1 + \dfrac{c_1}{P_1}}$, where $a_1, b_1, c_1 > 0$. When $P_1 = \sqrt{\dfrac{c_1}{a_1}}$, the CAR reaches its maximum value. The CAR exhibits a trend of initially increasing and then decreasing as the pump power increases, as shown in Figure 4.7(a). The curve depicting the relationship between the CAR and the SPDC pump power aligns with the theoretical expectations.

Next, we consider the cascading process. In the SPDC process, the idler photons are directly detected, while the signal photons undergo frequency conversion through the cascading process and are then measured in coincidence with the idler photons. The

number of converted signal photons $N_{2CC}$ and the signal single-channel count $N_2'$ can be expressed as:

$$N_{2,CC} = \alpha_C(P_2, \alpha_2\eta_2 N_0, T_2) \tag{9}$$

$$N_2' = \alpha_C(P_2, \alpha_2\eta_2 N_0, T_2) + \beta_C(P_2) + N_{nc}, \beta_C(P_2) + N_{nc} \gg \alpha_C(P_2, \alpha_2\eta_2 N_0, T_2) \tag{10}$$

Here, $P_2$ denotes the pump power of the cascaded process, T represents the crystal temperature, $\beta_C$ signifies the noise photons generated by the Raman process of the pump light within the CPPLN crystal during the cascaded process, and $N_{n1,2,c}$ encompasses environmental noise and dark counts. Given that the three factors influencing the counting rate of signal photons are mutually independent, the impact of each factor $N_{2CC}$ can be individually expressed as follows when considered in isolation:

$$N_{2,CC}(P_1) = a_{C1}(P_2, T_2)P_1 \tag{11}$$

$$N_{2,CC}(P_2) = a_{C2}(P_1, T_2)P_2^n, n = 2 \tag{12}$$

$$N_{2,CC}(T_2) = a_{C3}(P_2, \alpha_2\eta_2 N_0)\sin c^4(T_2) \tag{13}$$

Next, under different initial conditions, we analyze the trends of coincidence counts, accidental coincidence counts, and the CAR separately:
(1) Varying the SPDC pump power: The photon yield of the SPDC process shows a linear relationship with the SPDC pump power. At the same time, since the number of effective signal photons after cascading is much smaller than the number of noise photons, the signal single-channel counts exhibit the following relationship:

$$\beta_C(P_2) + N_{nc} \gg \alpha_C(P_2, \alpha_2\eta_2 N_0, T_2) \tag{14}$$

Therefore, when increasing the power of the SPDC process, $N_2'$ is no significant change, and the CAR expression can be written as:

$$CAR_{C1} = \frac{\alpha_1\eta_1\alpha_2'\eta_2' N_{2,CC}(P_1)}{N_1 N_2' \Delta\tau} = \alpha_1\eta_1\alpha_2'\eta_2' a_{C1}(P_2, T_2)\frac{P_1}{a_{C1}P_1 + b_{C1}} \tag{15}$$

At low SPDC pump power, the CAR exhibits an approximately linear relationship with the pump power $P_1$, however, as the power increases, the CAR gradually approaches a constant value $\dfrac{1}{a_{C1}}$, which is consistent with the trend shown in Figure 4.8(a).

(2) Varying the pump power of the cascaded process: Since the first-order cascaded process corresponds to two second-order nonlinear processes (SFG+DFG), the conversion efficiency exhibits a quadratic relationship with the pump power, the $N_{2CC}$ can be written as:

$$N_{2,CC}(P_2) = a_{C2}(P_1,T_2)P_2^n, n=2 \tag{16}$$

Meanwhile, as the pump power increases, the single-channel counts are primarily influenced by the Raman noise generated by the 1064nm pump light in the CPPLN crystal, leading to the accidental coincidence counts exhibiting the following relationship:

$$N_2' \propto \beta_C(P_2) \propto P_2^m, m=1 \tag{17}$$

Therefore, the CAR expression can be written as:

$$CAR_{C2} = \alpha_1\eta_1\alpha_2'\eta_2'a_{C1}(P_1,T_2)\dfrac{P_2^n}{a_{C2}P_2^m + b_{C2}} = \alpha_1\eta_1\alpha_2'\eta_2'a_{C1}(P_1,T_2)\dfrac{1}{a_{C2}P_2^{m-n} + b_{C2}P_2^{-n}} \tag{18}$$

By taking the first derivative of the expression $a_{C2}P_2^{m-n} + b_{C2}P_2^{-n}$ with respect to the power $P_2$, the following relationship can be obtained:

$$(m-n)a_{C2}P_2^{m-n-1} - nb_{C2}P_2^{-n-1} < 0 \tag{18}$$

This indicates that the CAR will increase with the rise in cascaded pump power. Furthermore, by taking the second derivative, the following relationship can be derived:

$$(m-n)(m-n-1)a_{C2}P_2^{m-n-2} + n(n+1)b_{C2}P_2^{-n-2} > 0 \tag{19}$$

From the equation (19), we can see that the rate of the CAR increases with cascaded pump power will gradually decelerate, which provides a clear explanation for the trend shown in Figure 4.8(b).

(3) Changing the crystal temperature: The operating temperature of the crystal primarily determines the phase-matching function of the second-order nonlinear process. As the temperature varies, the conversion efficiency will exhibit a trend dependence on the $\sin c^4$ function. In this process, the single-channel counts remain approximately constant, and the CAR can be expressed as:

$$CAR_{C3} = \frac{\alpha_1\eta_1\alpha_2'\eta_2' a_{C1}(P_2,\alpha_2\eta_2 N_0)}{N_1 N_2' \Delta\tau}\sin c^4(T_2) \qquad (20)$$

Therefore, as shown in Figure 4.8(c), the variation of the CAR with the temperature of the CPPLN crystal follows a $\sin c^4$ function relationship.